\documentclass[twocolumn,superscriptaddress,showpacs,amsmath,amssymb,aps,prb]{revtex4-1}
\usepackage{graphicx}
\usepackage[pdftex, colorlinks=true, linkcolor=blue, citecolor=blue, urlcolor=blue]{hyperref}
\usepackage{times}
\graphicspath{{./figures/}}
\DeclareGraphicsExtensions{.pdf}

\begin{document}
\title{A fast approach to Anderson localization for even-$N$ Dyson insulators}
\author{G. T. Stamatiou}
\author{S. N. Evangelou}
\affiliation {Department of Physics, University of Ioannina, Ioannina 45110, Greece}
\date{\today}

\begin{abstract}

Dyson insulators with random hoppings in a lattice  approach localization faster compared to the usual Anderson insulators with site disorder. For even-$N$ lattice sites the Dyson insulators mimic topological insulators with a pseudo-gap at the band center and the energy-level statistics obtained via the $P(S)$ distribution is of an intermediate type close to the Anderson localized Poisson limit. For odd-$N$ level-repulsion and Wigner statistics appears as in the quasi-metallic regime of $2D$ Anderson insulators, plus a single $E=0$ mode  protected by chiral symmetry. The distribution of the participation ratio and the multifractal dimensions of the midband state are computed. In $1D$ the Dyson state is localized and in $2D$ is fractal. Our results might be relevant for recent experimental studies of chiral localization in photonic waveguide arrays.  

\end{abstract}

\pacs{73.20.Fz, 71.55.Jv, 71.23.-k}
\maketitle

\section{Introduction}

The divergence of the density of states at the band center found by Dyson \cite{1} is probably the oldest result in the theory of disordered systems. It was followed by the pioneering paper of Anderson \cite{2} which showed localization via destructive interference of electron-waves due to disorder. In a tight-binding lattice description Anderson insulators have random on-site potentials and Dyson insulators have random hopping amplitudes between lattice sites \cite{3}. In the one-electron framework these two types of disorder better known as diagonal and off-diagonal can cause Anderson localization. Localization occurs either for strong disorder or low dimensionality which inhibit propagation (see also the recent activity on many-body localization in the presence of interactions between the electrons).\cite{4} The localized states usually show exponential spatial decay of the wave function measured by a finite localization length $\xi$. For chiral systems the exponential decay  is not guaranteed and other localization measures must be invoked, such as the energy level-statistics and the inverse participation ratio. For  Anderson insulators at the Fermi energy the corresponding conductance also decays. We shall show that for Dyson insulators localization occurs differently and can be better measured by means other than $\xi$. Our approach is to use numerical diagonalization  to get the spectra and the wave functions for finite chiral disordered chains and squares. The purpose of this paper is to show localization in finite chiral disordered systems and point out its differences from non-chiral ones.\cite{5,6,7,8}

The emergence of topological insulators and superconductors gave  systems with a gap also the four BdG universality classes for Bogolyubov quasi-particles.\cite{9} The prime examples of topological insulators are the quantum Hall due to a magnetic field and the spin quantum  Hall effect due to spin-orbit coupling.\cite{10} The states inside the gap are protected from localization at least for weak disorder. Our even-$N$ Dyson systems at the band centre have a pseudo-gap  and mimic topological insulators by showing  different even-odd $N$ behavior. The chiral symmetry is responsible which is absent for  the usual Anderson insulators. The chiral or sublattice symmetry is not destroyed by  random hoppings between the two sublattices of what is called a bipartite lattice.  As a consequence for any realization of disorder a Dyson insulator has an exactly  symmetric energy spectrum around the band center and for odd-$N$ a special $E=0$ mode. The corresponding Hamiltonian belongs to one of three chiral universality classes, the orthogonal for real symmetric matrices, the unitary in the presence of a magnetic field which breaks time-reversal symmetry and the symplectic  for systems with spin-orbit-coupling which break spin-rotation symmetry. The advantage for the studied chiral system the symmetry reduces the  Hamiltonian  into half since only the positive eigenergies are required, in other words instead of $H$ we study $H^{2}$ getting all the eigenvalues squared. We will show that the  $2D$ chiral disordered system can undergo a different kind metal-insulator transition driven by topologically induced Anderson localization.\cite{8}

In quasi-$1D$ chiral disordered systems the even-odd parity effect at $E=0$  for the localization length and the conductance was observed in \cite{5,6}.  For random hopping systems with even or odd number of coupled chains their  localization was different, the $1D$ Dyson singularity for the density of states $\rho(E)$ was found only for odd number of chains and for even number the usual exponential decay was recovered.  In other words, two decades ago the localization length of $2D$ chiral insulators was shown to depend on the even-odd parity for the number of chains in a wire different from Anderson insulators.

In this paper  via a numerical diagonalization of $H^{2}$ we demonstrate such even-odd $N$ effects also in the spectra and the corresponding eigenvectors of chiral disordered systems. For even-$N$ the Dyson insulators have a pseudo-gap at the band center and show a faster approach to localization.  For odd-$N$ the Dyson insulators behave as ordinary Anderson insulators with an additional unlike the rest mode at $E=0$. In fact, inside the $1D$ Dyson singularity the participation ratio  of the $E=0$ state  has shown it to be even more localized than the rest of states in the spectrum. In $2D$ the $E=0$ state is multifractal and to compute the exponents a size variation is required.

The approach to localization is shown via energy-level statistics near $E=0$.  For even-$N$ no weak localization corrections are expected\cite{3} and in $2D$ the obtained $P(S)$ is intermediate between Wigner and Poisson. For odd-$N$ Wigner statistics occurs for sizes shorter than the large localization lengths as for ordinary Anderson insulators.\cite{11} For large $N$ the statistics of strongly localised states is characterized by the Poisson distribution and the even-odd parity effect disappears.

In order to describe the spatial extend of the midgap mode its IPR  turns out to be a more suitable measure  for finite systems since $\xi$ is an asymptotic property for $N\to \infty$ when exponential localization occurs. In $1D$ a $\xi$ defined from exponential decay is dominated by fluctuations and is broadly distributed, apart from its average $\propto \ln^{2}(|E|)$  also it requires a typical length $\propto \ln(|E|)$.\cite{5} The $E=0$ state  described by $\xi$ contradicts what we find from the IPR which shows the opposite trend, the zero energy being the most localized state in the spectrum. In fact it has large fluctuations and is not extended due to the mild divergence of $\xi$. In $2D$ it was also impossible to describe the $E=0$ state by a single localization length\cite{12} and it turns out to be very sensitive to the boundaries of the system. We present the multifractal dimensions of this state. 

In chiral disordered systems large hoppings can connect  pairs of sites and dimerization takes place which corresponds to energies far from $E=0$.\cite{13} The energies close to $E=0$    arise from the remaining isolated sites which behave independently if connected. They  allow the accumulation of states responsible for the Dyson singularity as $E\to 0$. It was recently shown\cite{14} that $log$-terms in the hopping distribution can  lead to the absence of Dyson  singularity and to even stronger  localization known as super-localization. All these results suggest that instead of relying only on $\xi$ other measures must be also used in $2D$ to describe the peculiar delocalization of the $E=0$ state.

The wave phenomena in disordered media can be investigated for electromagnetic waves in complex structures, such as waveguide arrays which resemble a finite lattice. A wealth of recent experimental results \cite{15,16,17,18,19,20} combine disorder with non-linearity and it is possible to test various 
theories (e.g. topological effects, the critical exponents at the Anderson transition, 
etc.), including regimes that are difficult to access in disordered electronic 
systems. The ultracold atoms also show Anderson localization.\cite{21} The chiral problem can be studied in quantum spin-$1/2$ chains via the random exchange XX-spin model.\cite{22,23,24} There are four BdG universality classes which describe quasiparticles at a mean-field level which can be treated with disorder.\cite{10}

The main message of this paper is that hopping disorder is stronger than ordinary disorder and more easily leads to localization.\cite{25} In $2D$ for even-$N$ we find no Wigner statistics since the quasi-extended regime known for Anderson insulators is absent. The reason for this unusual localization is the adopted disorder if made strong also increases hopping which favours propagation which is the opposite of localization. The other intricacy of odd-$N$ Dyson insulators is the presence of the $E=0$ mode  which happens to be exactly at  half filling and its broadly distributed conductance shows an average which indicates inverse square root algebraic localization.\cite{12}

We study the orthogonal BDI chiral universality class where time-reversal invariance holds and the studied random matrix is real symmetric. In Sec. II we set up the Hamiltonian  for our numerical diagonalization approach in $N$-site  $1D$ and $2D$ chiral disordered systems. In Sec. III we present our results for the statistics of the first level near the band center for even-$N$ and odd-$N$ which are shown to be different. The inverse participation ratio distributions and the  multifractal exponents of the $E=0$ mode can be also found. In Sec. IV we discuss our results in relation to experiments before we present our conclusions in  Sec. V.

\section{The chiral Hamiltonian}
\label{sec:The chiral Hamiltonian}

We study Anderson localization of non-interacting particles in the phase-coherent quantum 
system of a bipartite lattice having real random hopping $t$. This off-diagonal disorder is achieved by distributing the hopping $t$ and also $\ln (t)$ from a box distribution of width $W$. The lattice consist of a total $N=L^{d}$ sites, $d$ is its dimension and the spinless Hamiltonian reads
\begin{equation}
 H =  \sum_{<i,j>}  \left( t_{i,j} c^+_{i}c_{j} +\text{hc} \right),   
\end{equation}  
where $c^+_{i}$ ($c_{i}$) creates (annihilates) a particle on site 
$i=1,2,\dots,N=L^{d}$ and $ < i,j >$ denote nearest-neighbour bonds in the lattice. We assume only nearest-neighbour hopping and no boundary conditions were imposed. We are careful since boundary conditions can break chiral symmetry. The time-reversal symmetry  holds and the 
$t_{i,j}$ connecting the two sublattices are real independent random variables drawn from a uniform distribution within $[-W/2;W/2]$ or  from the same distribution for the logarithm $\ln t_{i,j}$ which implies only positive 
$t_{i,j}$ in $[e^{-W/2}, e^{+W/2}]$. The first type of disorder, the distribution of $t_{i,j}$ includes negative hopping and gives even stronger disorder than the logaritmmic for the same $W$. The disorder which allows fluctuations  in the hopping signs is single valued and cannot be varied.\cite{26} A peculiar form of chiral disorder occurs for randomly placed vacancies where the chiral systems have a binary type disorder.\cite{8}

The chiral symmetry in the  $A, B$ sublattice basis is expressed by $\sigma_{z} H \sigma_{z}= -H$, where $\sigma_{z}$ is the third Pauli matrix which flips the sign of the wave function on one sublattice. The chiral matrix Hamiltonian $H$ can be decomposed into the off-block diagonal form
\begin{eqnarray}
H = \left( \begin{array}{ccc}
0 & H_{AB}    \\
H_{AB}^{+}  & 0 
\end{array} \right) \, 
\end{eqnarray}
with the nearest-neighbour sites $ < i,j  >$ belonging to the two interconnected A,B sublattices, respectively. The chiral symmetry and the corresponding squared matrix is
\begin{eqnarray}
H^{2} = \left( \begin{array}{ccc}
H_{AB}  H_{AB}^{+}  & 0 \\
 0 &  H_{AB}^{+} H_{AB}
\end{array} \right), \, 
\end{eqnarray}
where the random matrix $H_{AB}$ connects $A$, $B$.  In order to obtain the positive energy spectrum of $H$ it suffices to diagonalize only the block matrix product $H_{AB} H_{AB}^{+}$ and take the positive square root of the obtained eigenvalues. In the rest we study localization of $H$ by increasing the matrix size $N$. For the three chiral classes BDI, CII, AII the spectrum is symmetric and the eigenvalues always appear in pairs. Our study is limited to the BDI class.

The prime measure of localization is the localization length $\xi$ which is obtained from the exponential decay of the wave function. For the wave function component $\psi(r)$ on site $r$, e.g. in $1D$ 
\begin{equation}
\xi^{-1} =  -\lim_{r\to \infty} {\frac {1}{r}}\ln|\psi(r)|   
\end{equation}  
and $\xi$ is an asymptotic property. In $2D$ the situation is more involved and one has many $\xi$'s equal to the number of propagating channels. The definition of Eq. 4 ignores the main part of the wave function and can miss non-exponential decay (e.g. sub-localization or super-localization). Other less direct measures of localization include the distribution of the energy level-spacings $P(S) = P(E_{j}-E_{j-1})$ between the energy levels $E_{j}, j=1,2,...$ and for each state corresponding to $E_{j}$ the sum $IPR =  \sum_{r} |\psi(r)|^{4}$ defines the inverse participation ratio,  where $\psi(r)$ is the wave function component on site $r$. The generalization of IPR to the $q$ - th moment involve $|\psi(r)|^{2q}$ whose scaling defines the multifractal exponents $D_{q}$ for every $q$.

\section{Results}
\label{sec:Results}

\subsection{The level-statistics}
\label{sec:The level-statistic}

The distribution of the spacings between the energy levels 
\begin{equation}
P(S) = P(E_{j}-E_{j-1})
\end{equation} 
is well-known to flow from Wigner $P(S)\sim S exp(-S^2)$ for delocalized states to Poisson $P(S)=\exp(-S)$ for localized. In between the Anderson delocalization-localization transition  occurs and the critical distribution is scale-invariant. For even-$N$ the $P(S)$ gives the distribution of the pseudo-gap at the band center and for odd-$N$ it gives the distribution $P(E_{1})$ of the first level other than zero.

In $1D$ chiral systems both for even and odd-$N$ the obtained level-statistics rapidly approach the Poisson curve (Fig. 1(a), 2(a)) which implies localization as expected. In $2D$ for even-$N$ the absence of level-repulsion is observed (Fig. 1(b)) and localization occurs since the distribution for the gap  moves towards the Poisson curve.  For odd-$N$ instead for $S=E_1$ the  Wigner statistics is seen (Fig. 2(b))  which shows level-repulsion. In $2D$ for weak disorder the Wigner distribution is found as for  weak disorder Anderson insulators when the system behaves as if quasi-metallic.\cite{11} In $2D$ for even-$N=4,16,36$ and log-disorder with $W=.5, 1, 2$  the energy-levels give a $P(S)$ which rapidly approaches the Poisson curve. For odd-$N=9,25,49$ and the same disorder Wigner is found instead as for $2D$ Anderson quasi-metallic systems. 

\begin{figure}
\centerline{\includegraphics[width=\linewidth]{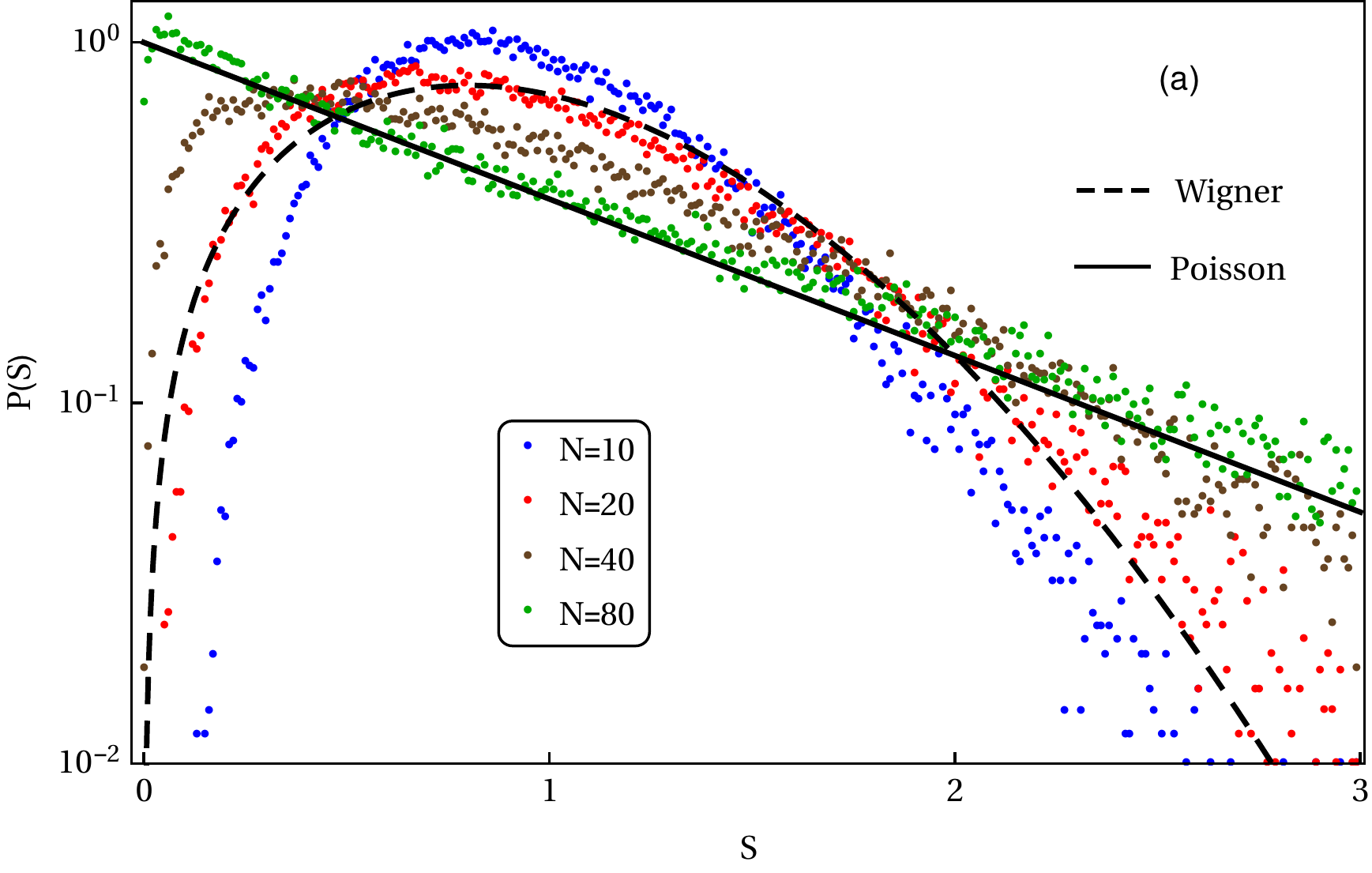}}
\centerline{\includegraphics[width=\linewidth]{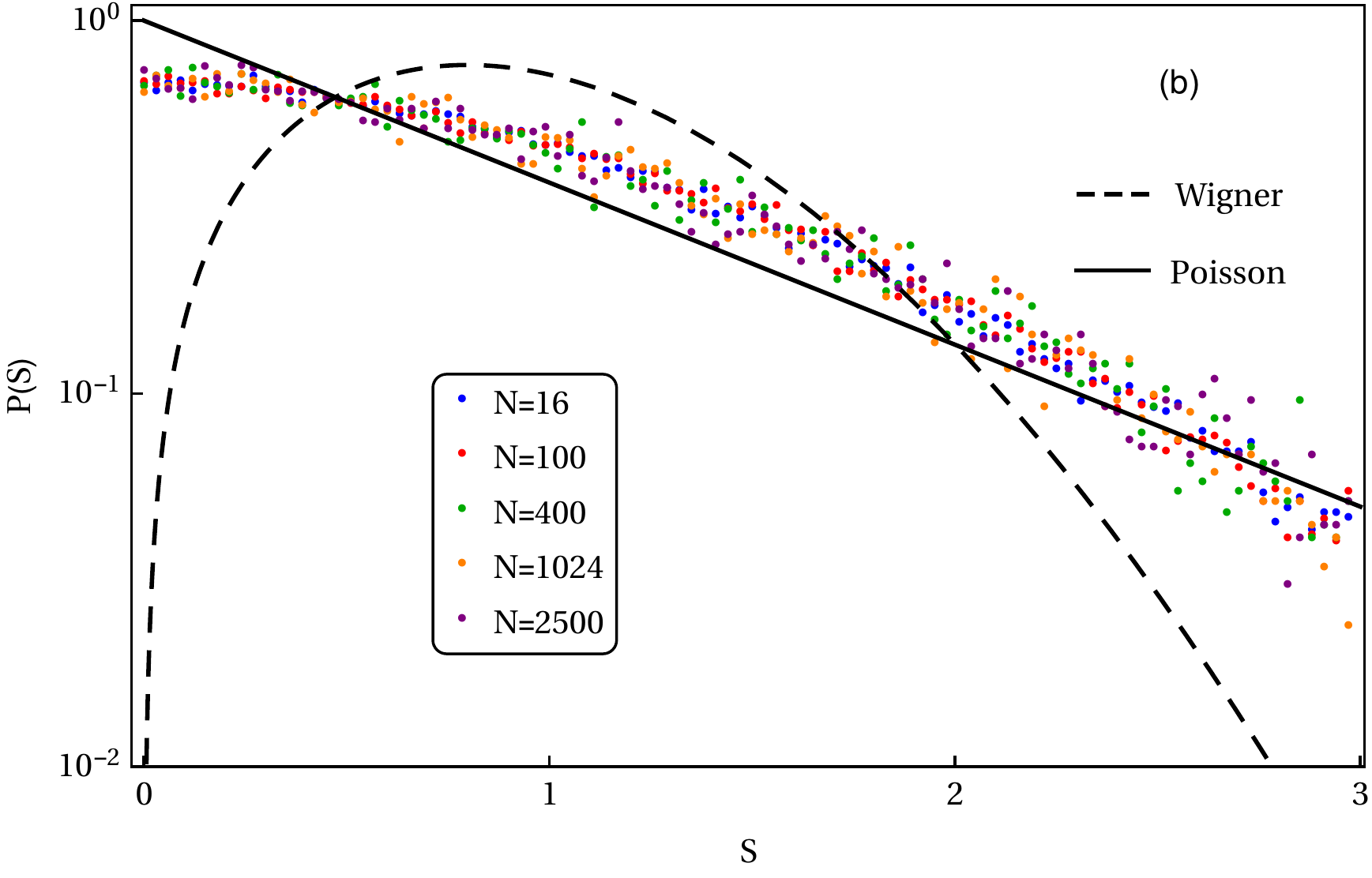}}
\caption{\label{fig:sketch}
The even-$N$ gap distribution at the band center in (a) $1D$  chains and (b) $2D$ squares. The hopping disorder has strength $W=1$ and  various sizes $N=L$ and  $N=L\times L$ are taken. The dashed line corresponds to Wigner statistics for extended states and the continuous line to the Poisson statistics for localized states. In $1D$ the data reach the Poisson (continuous line) which implies localization. In $2D$ as $N$ increases depending on the value of disorder  the obtained distribution is close to Poisson. The data actually land on an intermediate semi-Poisson curve which is linear for small $S$ Wigner-like and has a  simple exponential Poisson tail for large $S$. In this case level-repulsion and the Gaussian Wigner-decay for the tail is not found. We chose $80000$ realizations in $1D$ and $30000$ in $2D$ with fixed boundary conditions.}
\end{figure}

\begin{figure}
\centerline{\includegraphics[width=\linewidth]{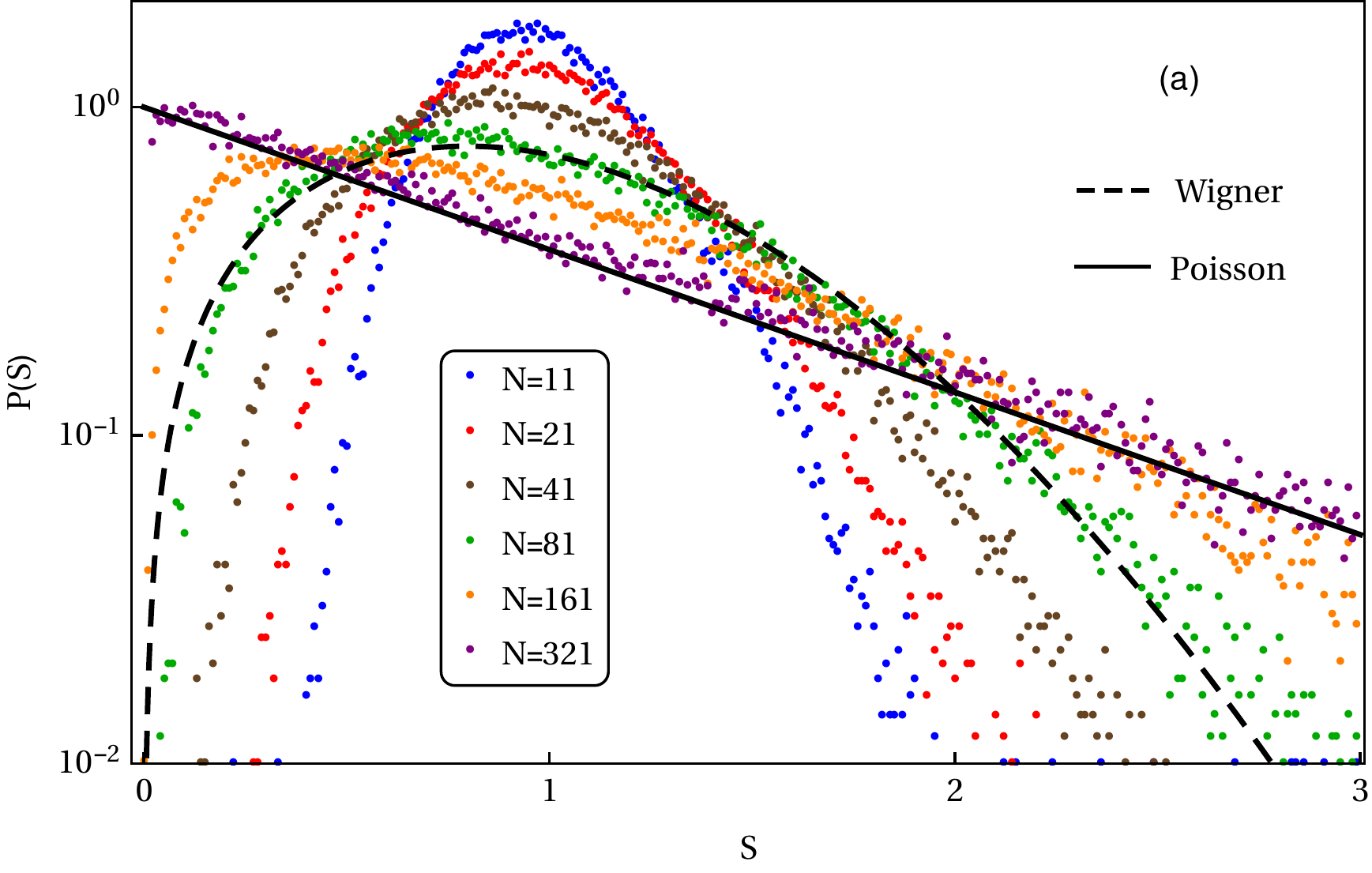}}
\centerline{\includegraphics[width=\linewidth]{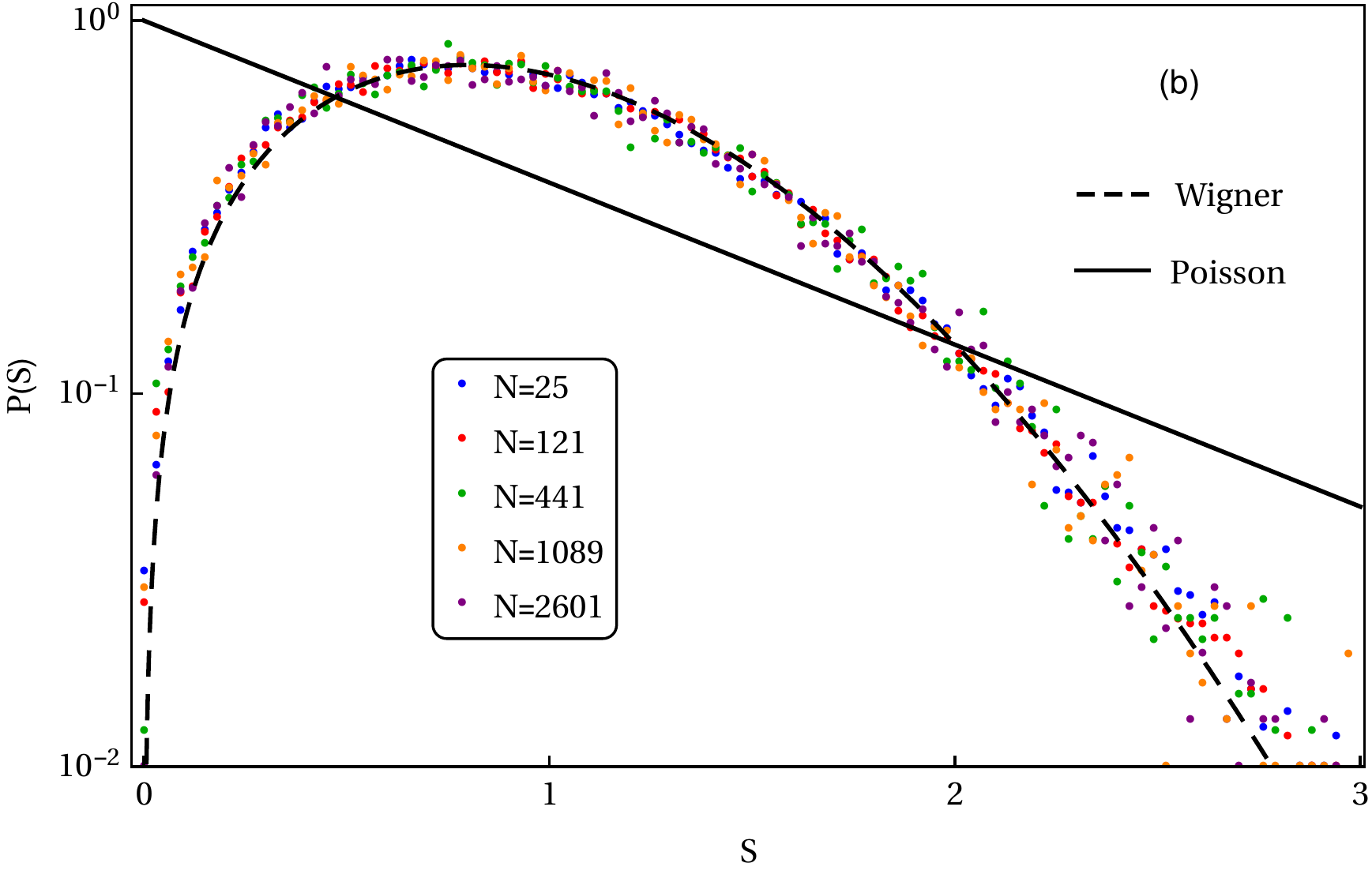}}
\caption{\label{fig:sketch}
The odd-$N$ distribution of the first positive energy level in (a) $1D$ chains and (b) $2D$ squares with the same hopping disorder of strength $W=1$ and various sizes $N=L$, $N=L\times L$, respectively. The $E=0$ mode is also present. The dashed line corresponds to Wigner statistics for extended states known in the quasi-metallic $2D$  regime for sizes smaller than the localization lengths and the continuous line is the Poisson statistics for localized states which can be reached for strong disorder. For the chosen sizes odd-$N$ the statistics of $2D$ chiral systems  is Wigner as in the quasi-metallic regime of Anderson insulators. In the case of odd-$N$ the Poisson distribution is not seen. }
\end{figure}

\subsection{The participation ratio}
\label{sec:The distribution of the participation ratio}

For each energy level $E_{j}$ is the sum over all sites 
\begin{equation}
IPR =  \sum_{r} |\psi(r)|^{4}   
\end{equation} 
gives the inverse number of sites where the wave function has a significant amplitude. The IPR of the $E=0$ state concerns averages obtained by  creating a random ensemble repeating the diagonalization many times and shows localization as $N$ increases. The other case is to take a single but very large random sample usually done in the calculation of $\xi$. In $1D$ for the $E=0$ state $\ln(|\psi|)$ executes a simple random walk and also in $2D$ its statistics for the sizes taken is Gaussian.

In $1D$  the distribution $P(\ln(IPR))$ for the localized $E=0$ state is presented in Fig. 3(a),(b). In $1D$ the $E=0$ state is not extended as the divergent $\xi$ might suggest. In Fig. 3(a) inset the distribution of $PR=1/IPR$ for log chiral disorder of $W=1$  despite its large fluctuations converges for large $N$ and its averaged PR peaks at lengths below twenty. The $E=0$ state is strongly localized. In $2D$ the $P(\ln(IPR))$ does not converge with increasing $N$ and the IPR maxima scale with $N$ (Fig. 3(b)) which allows to define the fractal dimensions, e.g. $D_{2}$.\cite{27,28} The $E=0$ wave function probability has a chessboard pattern density instead of a high amplitude followed by an exponential decay seen in Anderson insulators. This can be attributed to the choice of the random hopping distribution which involves both positive and negative hopping terms, that is hopping sign fluctuations present additional disorder in the chiral system. In $1D$ for $N=3$ the $E=0$ value of $IPR={\frac {t_{1}^{4}+t_{2}^{4}}{(t_{1}^{2}+t_{2}^{2})^{2}}}$ is higher than the non-zero state $IPR={\frac {t_{1}^{4}+t_{2}^{4}+t_{1}^{2}t_{2}^{2}}{2(t_{1}^{2}+t_{2}^{2})^{2}}}$ and the $E=0$ Dyson state surprising appears even more localized than the rest of the spectrum. This is observed for  odd-$N$ also in $2D$ but in order to characterize its localization one needs a size variation.

\begin{figure}
\centerline{\includegraphics[width=\linewidth]{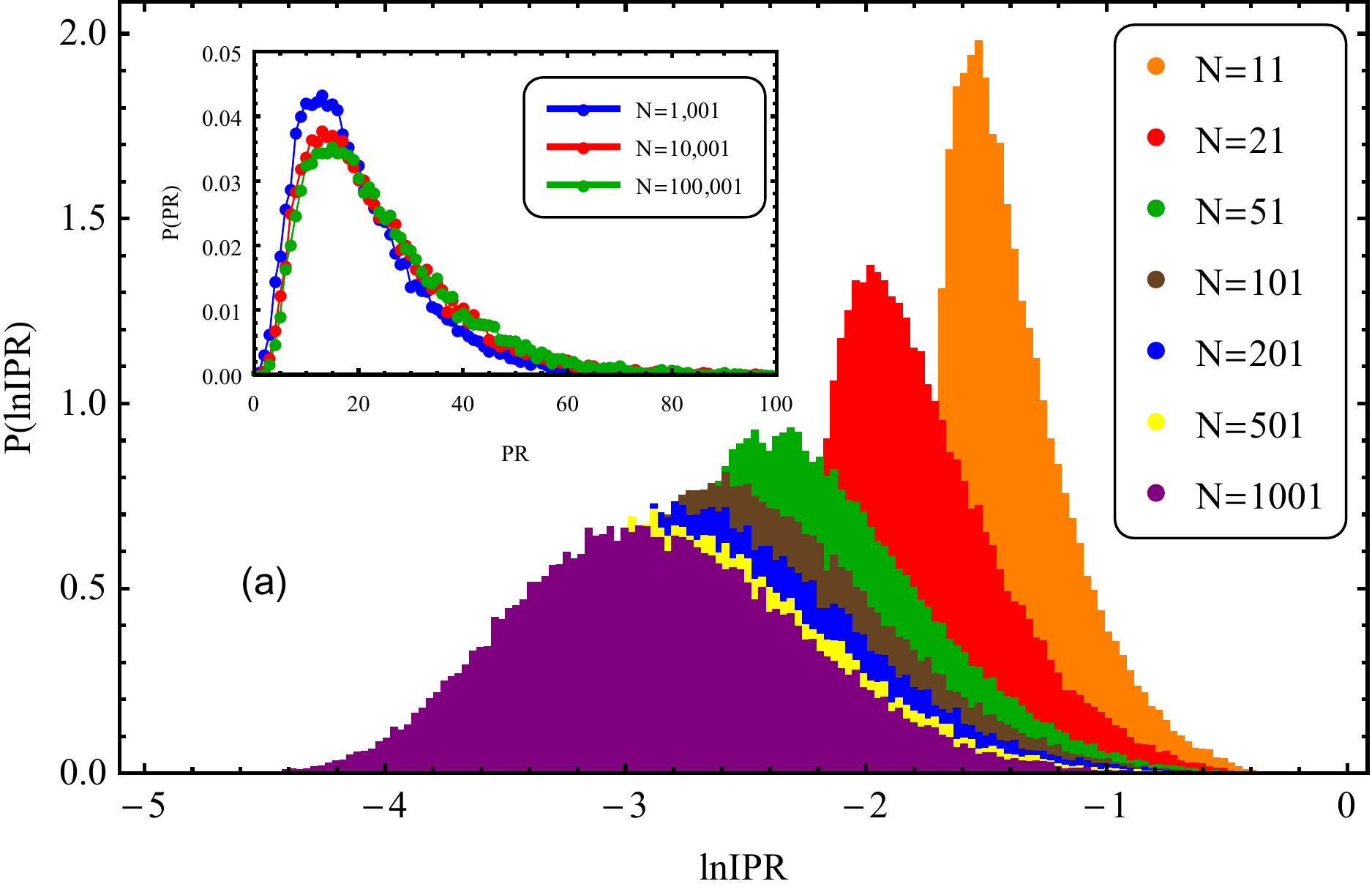}}
\centerline{\includegraphics[width=\linewidth]{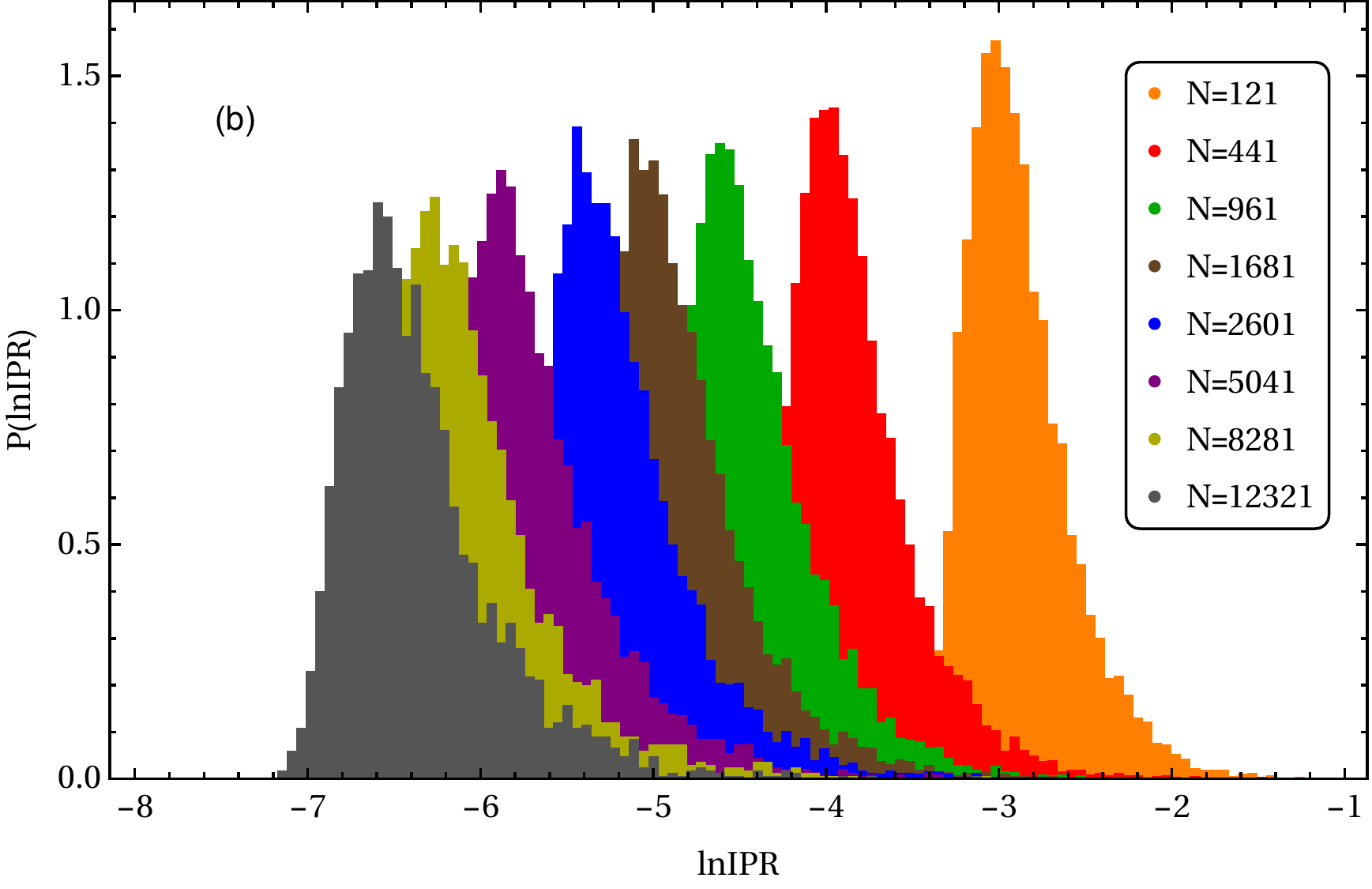}} 
\caption{\label{fig:sketch}
a) In $1D$ for the log-disordered chiral system with $W=1$ the distribution for the log inverse participation ration (IPR) of the $E=0$ state is shown. Inset: The convergence of PR of the $E=0$ state for log disorder of $W=1$. (b) In 2D the fluctuations of IPR grow by increasing $N$, IPR does not converge and its scaling  with size gives the multifractal dimensions.}
\end{figure}
 
\subsection{The fractal dimensions}
\label{sec:The fractal dimensions}

The generalization of IPR to the $q$ - th moment involves $|\psi(r)|^{2q}$ whose scaling with the linear size, in $2D$ where $L= N^{1/2}$  define the multifractal exponents $D_{q}$. There are two ways to proceed, one can get either the averaged IPR or the averaged $\ln(IPR)$. The scaling with size 
\begin{equation}
\overline {\sum_{r} |\psi(r)|^{2q}}\sim L^{-D_{q}} 
\end{equation}  
and
\begin{equation}
exp(\overline {\ln(\sum_{r} |\psi(r)|^{2q})})\sim L^{-\widetilde D_{q}} 
\end{equation}  
defines the dimension $D_{q}$ and the typical dimension $\widetilde D_{q}$. For $q=0$ the $D_{0}$ is the space dimension and for $q=2$ it gives $D_{2}$. The $2D$ averaged and typical fractal dimensions are plotted in Fig. 4. A freezing transition of exponents occurs at the $q_{c}$ where $D_{q}$ and $\widetilde D_{q}$ begin to differ, for $q>q_{c}$ non-typical events dominate the average.

For logarithmic disorder of $W=1$ we find $D_{q}= D_{0}-c q^{\alpha}$, for $q<<1$, $D_{0}=2$, $c=const.$ and $\alpha\simeq 0.7$. The Legendre transform of  $D_{q}$ gives the $f(\alpha)$ spectrum which shows a maximum at $\alpha_0=2.59$.

\begin{figure}
\centerline{\includegraphics[width=\linewidth]{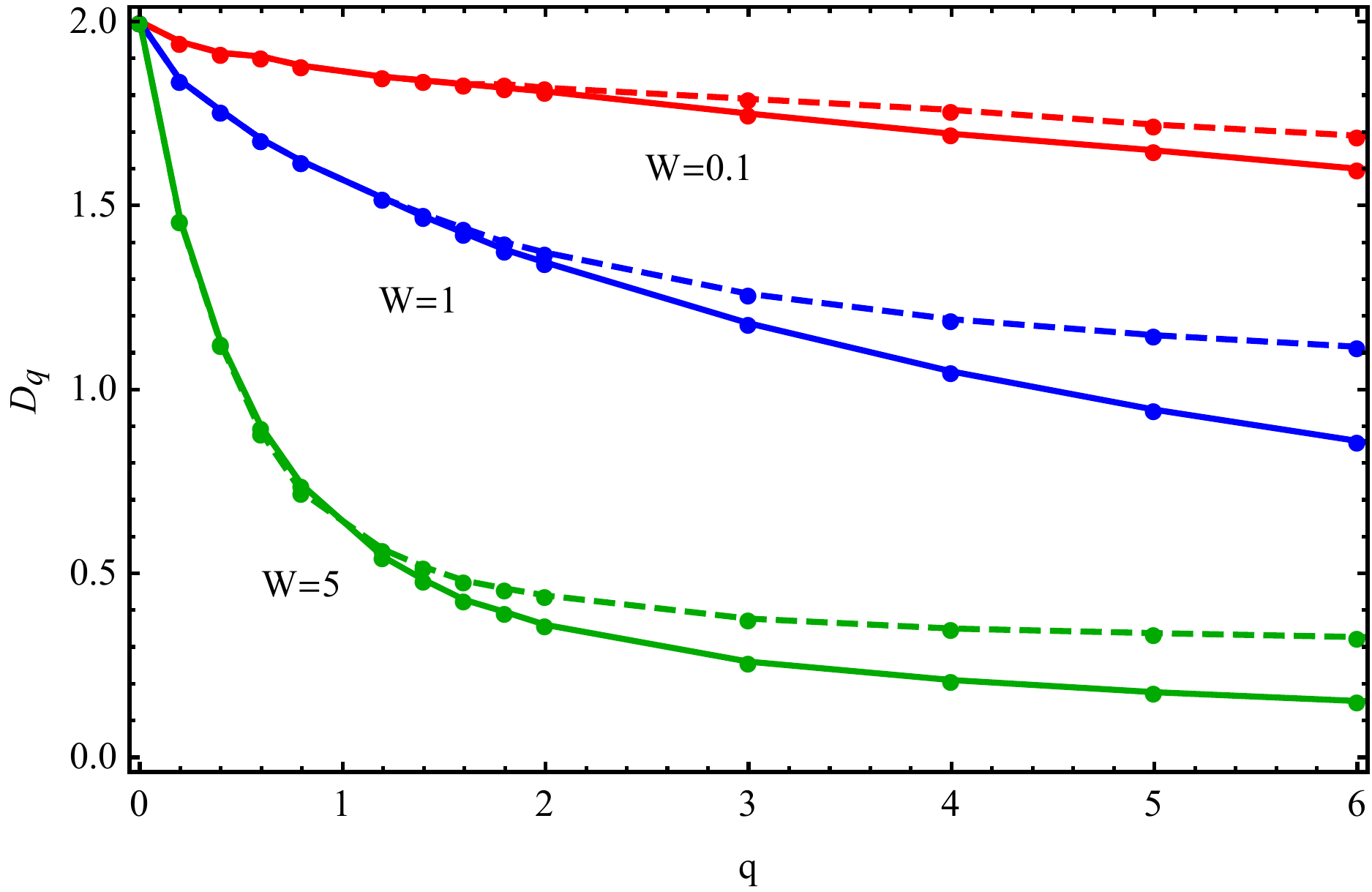}}
\caption{\label{fig:sketch}
The 2D generalized dimensions $D_{q}$ associated with the averaged values (full lines) and $\widetilde D_{q}$ with the typical values (dotted lines). When the non-typical events begin to dominate the two lines separate and for large $q$ the mean values are dominated by non-typical events.}
\end{figure}

\section{Discussion}
\label{sec:discussion}

Chiral systems with hopping disorder are less studied than Anderson systems with random site potentials. Large hopping implies higher propagation and in the case of random hoppings both propagation and localization occur. The disorder between sites  can be achieved by randomization of the distances between regular lattice sites. The chiral systems make up three chiral universality classes which can be studied via the eigensolutions of $H^{2}$. In our case of real hoppings time-reversal invariance is also preserved and due to chiral symmetry the spectrum is symmetric around the band center. The insulating behaviour in $1D$ and $2D$ is supposed to occur for all kinds of non-correlated disorder and at the lower critical dimension for localization unless time-reversal or spin-rotation symmetry is broken.  The main novelty of a chiral system is the Dyson singularity for the accumulation of the density of states $\rho(E)\sim {\frac {1}{E}}$ as $E\to 0$. The other is the even-odd parity effect for finite number of sites $N$.  In a chiral system as the number of sites $N$ grows  the states are more localized for even-$N$ than for odd-$N$ where the Dyson insulators behave as Anderson insulators with an extra $E=0$ mode. In $1D$ the $E=0$ mode is found strongly localized and in $2D$ it is multifractal.

This work was partially motivated by recent experiments on optical arrays. \cite{15,16,17,18,19,20} 
In these works light propagates along the waveguides in $z$-direction and Anderson localization in 
the transverse $x$--$y$ plane is also experimentally studied  by investigating the 
spreading of a local excitation to the neighboring waveguides. In such optical $2D$ systems light waves  through disordered photonic lattices can become localized due to multiple scattering effects. The density-density  Hanbury-Brown and Twiss density correlation measurement carry a distinctive signature for Dyson  or Anderson  type of disorder. The density measurements can distinguish systems with chiral symmetry.  The experiments which detect localization are also done with ultra cold atoms created by laser speckles.\cite{21} The chiral symmetry is also important for Graphene which is known to be a marginal topological insulator. Its ultra-thin layered structure and purity requirements in preparing the samples cannot avoid creating some disorder and the disordered graphene  was studied by transfer-matrix techniques in \cite{29}.  Two propagating channels exist and the $E=0$ mode is supposed to be resilient to localization.  For $L\times M$ graphene samples with positive and negative random hopping the conductivity is independent of the ratio $L/M$ and  for disorder with only positive values instead the conductance was independent of $L/M$.  In both cases  at the Dirac point  the conductivity is $4 e^{2}/h$.

\section{Conclusion}
\label{sec:conclusion}

We have studied $1D$ and $2D$ chiral disordered systems where Anderson localization is always expected. The chiral systems with hopping disorder thought to be an exception to this rule since the Dyson singularity of $\rho(E)$ shows accumulation of energy levels as $E\to 0$. The question whether chiral disordered systems approach localization  differently is addressed in this paper. Our results via the  diagonalization of $H^{2}$ show that Dyson insulators  even faster approach  Anderson localization and we confirm a topological even-odd-$N$  number of lattice sites effect known for the number of chains in wires. For even-$N$ a quasi-gap is found in the middle of the band and for odd-$N$ an $E=0$ midgap isolated  unusually localized state appears in $1D$ and in $2D$ it is multifractal. For even-$N$ faster approach to localization   via level-statistics close to the band center is shown since the obtained  intermediate $P(S)$ is closer to Poisson. For odd-$N$ level-repulsion  and in 2D Wigner statistics is found  as for Anderson insulators in the quasi-metallic regime. We also show the participation ratio and the multifractal dimensions of the midgap state. 

The even-odd $N$ parity effect in a chiral system is a sign of topology and in Dyson insulators  localization occurs  despite the extra chiral symmetry. Moreover it is  different for even and odd system sizes. For even-$N$  more localized states appear and we obtain an intermediate $P(S)$ which points towards localization. The obtained $P(S)$ for the whole band is independent from the choice of boundary conditions.\cite{26} For odd-$N$ Wigner is found as for usual $2D$ Anderson insulators and an $E=0$ mode  appears. For large sizes both even and odd-$N$ approach the localized Poisson limit. The Dyson insulators show different localization properties near $E=0$,  that is a chessboard pattern for the distribution of the amplitude of $\psi$ compared to Anderson insulators where for localized states the amplitude is distributed in a single area. Other universality classes, e.g. the symplectic case where the Dyson singularity does not appear and is believed to show a level-repulsion effect.\cite{5} The $E=0$ Majorana modes  similarly can be studied in the D, DIII universality classes. Quantum walks can be studied in chiral chains to see if they show long-range correlations.\cite{30,31}

Our results  are relevant for recent experimental studies of localization in photonic waveguide arrays.  In linear optical systems light scattering into disordered waveguide arrays the inter-spacings between guides  randomly changes and the nature of disorder is chiral. This chiral  symmetry is  inherent  and the distribution of photon numbers becomes supethermal. The probability that no photon will be detected becomes the highest and the sub-thermal light is largely inaccessible.\cite{15}  Chiral systems can also appear with diagonal disorder engineered by having a random potential equal between positive and negative sites.\cite{32} In higher dimensions the Bethe-lattice of $K+1$ neighbours also shows a Dyson singularity for $K<4$ which vanishes for higher $K$ while for $K\to \infty$ the Wigner semi-circle law for the $\rho(E)$ appears.\cite{33} The density of states near the band center is estimated as $\rho(E)\approx \rho(0)-a |E|^{\frac{K-1}{2}}$, $a=const$ as $E\to 0$. It would be interesting to see if an even-odd effect exists  at the $3D$ Anderson transition where the same localization length exponent $\nu=1.3$ exists and a lower multifractal dimension $D_{2}=0.69$\cite{34} instead of $D_{2}=1.57$ for the usual $3D$ Anderson transition.

In summary, our study has shown even-odd $N$ effects in the level statistics of chiral disordered systems.
For even-$N$ localized states with the Poisson statistics which appears earlier than for odd-$N$ where an $E=0$ also exists.  Our numerical study for odd-$N$ rules out  the presence of Wigner. In Dyson insulators chiral disorder is never weak since the hopping also affects propagation. The $E=0$ Dyson mode is more localized than the rest of the energies in the system and the usual localization length fails to measure the spatial extend of  this state. In conclusion, for Dyson insulators we have shown that the parity of the site number plays an important role. Our even-odd study removes certain doubts about numerical results pointed out in the field theoretic non-linear sigma model studies predicted what is called topological kind of transition in $2D$. \cite{8} Our numerical findings via the participation ratio show that Dyson insulators approach a different kind of localization. It is also important to find out what happens in the presence of strong interactions.\cite{35} For even-$N$ there is no quasi-metalic regime and no weak- localization corrections. For odd-$N$ things are the same as Anderson insulators with an extra mode at $E=0$. In chiral systems the disorder is never too weak and the localization length $\xi$ does not tell the full story. The divergent density of states found by Dyson is not sufficient to show delocalization at $E=0$.\cite{36}


\begin{thebibliography}{99}

\bibitem{1} 
F. J. Dyson, 
Phys. Rev. \textbf{92}, 1331 (1953). See also
\textit{50 years of Anderson localization}, World Scientific (2010).

\bibitem{2} P. W. Anderson, 
Phys. Rev. B \textbf{109}, 1492 (1958).

\bibitem{3} 
R. Gade  and F. Wegner. Nucl. Phys. B360 (1991). In  $2D$ chiral symmetry classes BDI, AIII, CII  quantum interference effects are absent to all orders of perturbation theory. See also R. Gade, Nucl. Phys. B398, 499 (1993).

\bibitem{4}
D.J. Luitz, N. Laflorencie and F. Alet,
Phys. Rev. B \textbf{91}, 081103 (2015)

\bibitem{5} 
P. W. Brouwer, A. Furusaki, C. Mudry and S. Ryu,
arXiv:cond-mat/0511622

\bibitem{6} 
C. Mudry, P. W. Brouwer, and A. Furusaki, Phys. Rev. B 62, 8249 (2000).
In the diffusive regime of chiral disordered systems there is no weak-localization correction to the conductance and exponential localization occurs only for even number of transmission channels in which case the localization length does not depend on whether time-reversal and spin-rotation symmetry are present or not. For odd number of channels the  conductance decays algebraically.

\bibitem{7} E. J. K\"{o}nig, P. M. Ostrovsky, I. V. Protopopov, and A. D. Mirlin
Phys. Rev. B 85, 195130 (2012) for random hopping models on $2D$ bipartite lattices at the band center have shown quantum localization effects as nonperturbative contributions controlled by topological vortexlike excitations of the sigma model.  
 
\bibitem{8} 
P. M. Ostrovsky, I. V. Protopopov, E. J. K\"{o}nig, I. V. Gornyi, A. D. Mirlin, and M. A. Skvortsov,
Phys. Rev. Lett. 113, 186803 (2014) studied $\rho(E)$ in $2D$ with vacancies.

\bibitem{9} 
A. Atland and M. R. Zirnbauer, PRB \textbf{55}, 1142 (1997). 

\bibitem{10} 
M. Z. Hasan and C. L. Kane
Rev. Mod. Phys 82, 3045 (2010).

\bibitem{11} V. I. Falko and K. B. Efetov, Phys. Rev. B \textbf{52}, 17413 (1995) in $2D$ disordered systems for sizes below the localization length find generic multifractal wave functions.

\bibitem{12} 
 P. W. Brouwer, E. Racine, A. Furiouser, Y. Hatsugai, Y. Morita, and C. Mudry, 
Phys. Rev. B \textbf{66}, 014204 (2002).


\bibitem{13} 
O. Motrunich, K. Damle, D. A. Huse 
Phys. Rev. B \textbf{65}, 004206 (2002).

\bibitem{14} 
A. Krishna and S. Bhatt,  
arXiv:2004.000064

\bibitem{15} 
A. Szameit,  
Nature Physics \textbf{11}, 895–896(2015).

\bibitem{16} 
H. E. Kondakci, A. F. Abouraddy, B. E. A. Saleh 
Nature Physics \textbf{11}, 930–935(2015).

\bibitem{17} 
H. E. Kondakci, A. F. Abouraddy, B. E. A. Saleh 
Scientific Reports volume 7, Article number: 8948 (2017) 

\bibitem{18} 
L. Martin, G. Di Giuseppe, A. Perez-Leija, R. Keil, F. Dreisow, M. Heinrich, S. Nolte, 
A. Szameit, A. F. Abouraddy, D. N. Christodoulides, and B. E. A. Saleh, 
Optics Express \textbf{19}, 13636 (2011). 


\bibitem{19} 
Y. Lahini, Y. Bromberg, Y. Shechtman, A. Szameit, D. N. Christodoulides, R. Morandotti, and Y. Silberberg,
Phys. Rev. A \textbf{84(R)}, 041806 (2011). Hanbury Brow and Twiss correlations.

\bibitem{20} S. Stützer, Y. V. Kartashov, V. A. Vysloukh, A. Tünnermann, S. Nolte, M. Lewenstein, L. Torner, and A. Szameit, Optics Letters 37,  1715-1717 (2012) 


\bibitem{21} D.H. White et al. arXiv: 1911.0458
The $2D$ Anderson systems for weak disorder have too large localization lengths and this is the reason of the Wigner statistics in $2D$. The localized regime requires very large systems to be reached. 
 
\bibitem{22} 
S. Sachdev,  {\cal {Quantum Phase Transitions}}, Cambridge University Press (2011).

\bibitem{23} 
D. Baeriswyl and G. Ferraz,  arXiv 1207.65593

\bibitem{24} 
L. Balents and M. P. A. Fisher, Phys. Rev. B \textbf{56}, 12970 (1997)

\bibitem{25} 
M. Hjort and S. Stafstr\"{o}m,  
Phys. Rev. B \textbf{63}, 113406 (2001).

\bibitem{26}
S. N. Evangelou and D. E. Katsanos, J. Phys. A: Math. Gen. 323 (2003)

\bibitem{27}
M. Inui, S. A. Trugman and E. Abrahams, 
Phys. Rev. B. \textbf{49}, 3190 (1994).

\bibitem{28} 
S.-J Xiong and S.N. Evangelou, Phys. Rev. B.  \textbf{64}, 113107 (2001).


\bibitem{29} 
S.-J Xiong and Ye Xiong, PRB \textbf{76}, 214204 (2007).

\bibitem{30} 
 Q. Zhao and J. Gong,
Phys. Rev. B \textbf{92}, 214205 (2015).


\bibitem{31} 
 H. Obuse and N. Kawakami, 
Phys. Rev. B \textbf{84}, 195139 (2011).

\bibitem{32} V. K. Varma, S. Pilati, and V. E. Kravtsov
Phys. Rev. B \textbf{94}, 214204 (2016).

\bibitem{33} 
V. Bapst and G. Semerjian, J Stat Phys, 51 (2011).

\bibitem{34} A. M. Garcia-Garcia and E. Cuevas,
Phys. Rev. B 74, 113101 (2006).

\bibitem{35} 
C. P. Chen, M. Szyniszewski and H. Schomerus,  
arXiv 2000.0753  (2020).


\bibitem{36} 
A. Comtet, C. Texier, Y. Tourigny, 
J. Phys. A: Math. Theor. \textbf{46}, 254003 (2013).  
The reason $\xi$ appears to over-estimate localization is attributed to the fact that real eigenstates should vanish at both ends of a chain instead of one and thus become more localized.






\end{thebibliography}
\end{document}